# All-optical switching in a continuously operated and strongly coupled atom-cavity system

Sourav Dutta* and S. A. Rangwala

*Raman Research Institute, C. V. Raman Avenue, Sadashivanagar, Bangalore 560080, India*

We experimentally demonstrate collective strong coupling, optical bi-stability (OB) and all-optical switching in a system consisting of ultracold $^{85}$Rb atoms, trapped in a dark magneto-optical trap (DMOT), coupled to an optical Fabry-Perot cavity. The strong coupling is established by measuring the vacuum Rabi splitting (VRS) of a weak on-axis probe beam. The dependence of VRS on the probe beam power is measured and bi-stability in the cavity transmission is observed. We demonstrate control over the transmission of the probe beam through the atom-cavity system using a free-space off-axis control beam and show that the cavity transmission can be switched on and off in micro-second timescales using micro-Watt control powers. The utility of the system as a tool for sensitive, in-situ and rapid measurements is envisaged.

Systems with atoms placed inside a cavity have been a subject of study for many decades [1–18]. Apart from fundamental physics [19], the motivation for studies with optical cavities lies in a large variety of applications in optical communication, quantum communication, quantum computing [20,21] and, as we suggest here, in sensitive measurement of interactions. Essential requirements for many of these applications are atom-cavity strong coupling [9,11,12,14,21] and all-optical switching of the cavity output light [7,10,22,23]. Perhaps the most important goal is to engineer all-optical switches that are fast, yet can be operated with minimal power [22–24]. To this end, significant progress has been made in cavity QED systems consisting of a single atom strongly coupled to a high finesse cavity [21], which however require extremely precise system control. Here we study the relatively less explored complementary system consisting of an ensemble of trapped ultracold atoms collectively coupled to a low finesse cavity [9–11]. This results in a significant technical simplification and ease with which a low intensity, fast all-optical switch can be implemented.

In this article, we show that atom-cavity collective strong coupling can be achieved on a non-cycling (i.e. open) transition in a continuously operated $^{85}$Rb dark-spot magneto-optical trap (DMOT) [25,26]. The signature of collective strong coupling is vacuum Rabi splitting (VRS) which is observed using a weak on-axis *probe* beam. The dependence of VRS on the *probe* beam power is measured and optical bi-stability (OB) in the cavity transmission is observed. Control over the nature of OB curve using a free-space off-axis *control* beam is demonstrated. We finally show that the cavity transmission can be switched on and off in micro-second timescales using micro-Watt *control* powers. Remarkably, a DMOT of ultracold atoms coupled to a cavity can be operated analogous to both strongly coupled atom-cavity systems as well as weakly coupled vapor cell based cavity systems, and retain advantages of the respective systems.

The details of the overall experimental apparatus which consists of an atom trap and an ion trap at the mode center of a low finesse optical cavity has been described elsewhere [11,27]. A schematic representation of parts relevant for the present experiments is shown in Fig. 1(a). The $^{85}$Rb DMOT is loaded from a Rb dispenser source. The DMOT is formed by three mutually orthogonal pairs of counter-propagating cooling beams and two mutually orthogonal repumping beams. The magnetic field gradient for the DMOT is ~22 Gauss/cm. The cooling and repumping lights are derived from two separate external cavity diode lasers (ECDLs). The cooling (repumping) beams are detuned by -12 MHz (+20 MHz) from the $F = 3 \rightarrow F' = 4$ ($F = 2 \rightarrow F' = 3$) atomic transition, are each 1 cm in diameter and each have 7 mW (2.2 mW) optical power. The repumping beams have their centers darkened with an opaque disc of 2 mm diameter such that no repumping light is present where ultracold atoms remain trapped – this pumps >95% of the ultracold $^{85}$Rb atoms to the ground non-fluorescing $F = 2$ state. The low fluorescence of the trapped atoms minimizes the otherwise deleterious effect of the fluorescent photons being coupled to the cavity. We measure the number of ultracold atoms in the DMOT by instantaneously turning it to a bright MOT using a repumping light tuned to the $F = 2 \rightarrow F' = 3$ transition and recording the fluorescence on a calibrated photomultiplier tube (PMT). We typically have ~$10^6$ atoms in the DMOT at a density of ~$10^{10}$ cm$^{-3}$.

The DMOT is positioned at the center of the Fabry-Perot cavity by monitoring (on a CCD camera) the fluorescence (of a bright MOT) that is out-coupled though the cavity (see Fig. 1(a)). The cavity consists of a pair of curved mirrors (radius of curvature 50 mm) separated by $L$ = 45.7 mm. The cavity waist for the TEM$_{00}$ mode is 78 µm and the finesse is measured to be ~ 650. One of the cavity mirrors is mounted on a piezoelectric stack that allows tuning the cavity length by a few µm. The length of the cavity is adjusted to be resonant with the $F = 2 \rightarrow F' = 3$ atomic transition (whose frequency is denoted by $\omega_{23}/2\pi$).

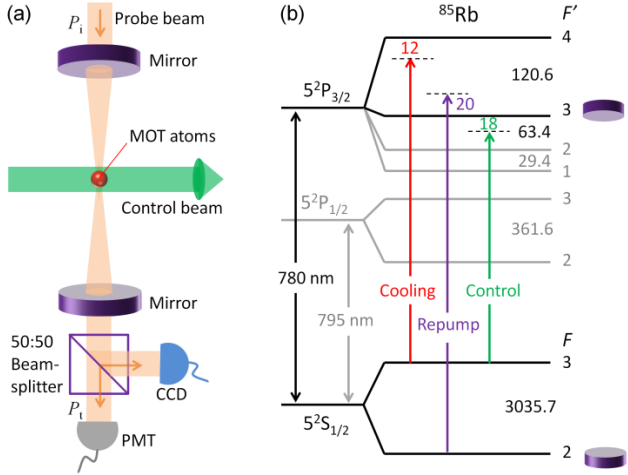

FIG. 1. (a) Schematic of the experimental setup. (b) Energy levels diagram of $^{85}$Rb. The black (grey) lines represent energy levels relevant (not relevant) for the present experiment. The frequency spacing between consecutive $F$ levels and the detuning of the lasers are marked (in MHz).

The atom-cavity system is probed with an on-axis *probe* beam whose frequency $\omega_p/2\pi$ is tuned and the transmission monitored on a PMT. Only a small fraction of the incident *probe* power ($P_i$) is coupled into the cavity due to imperfect mode-matching between the *probe* beam and the cavity mode (the cavity transmitivity is ~$10^{-5}$ of $P_i$). In what follows, we measure and present the incident *probe* power ($P_i$), and not the power coupled into the cavity. The *control* beam that we use to alter the atom-cavity coupling is incident along a direction perpendicular to the cavity axis (Fig. 1(a)). The frequency $\omega_c/2\pi$ of the *control* beam is detuned by -18 MHz from $\omega_{33}/2\pi$, the frequency of the $F = 3 \to F' = 3$ transition. The energy level diagram with all relevant frequencies is depicted in Fig. 1(b). The DMOT is operated continuously, i.e. cooling beams, repumping beams and the magnetic field are always on, which allows measurements to be performed continuously.

The interaction between a two-level atom (here, the $F = 2$ and $F' = 3$ levels of $^{85}$Rb) with a single mode of the electromagnetic field within a cavity leads to an alteration of the transmission though the cavity. The single atom-cavity coupling constant $g_0 = \sqrt{\mu_{23}^2 \omega_{23}/2\hbar\epsilon_0 V_c}$ determines the strength of the coupling, where $\mu_{23}$ is the transition dipole moment for the $F = 2 \to F' = 3$ transition, $\epsilon_0$ is the permittivity of free space and $V_c$ is the cavity mode volume. For our cavity parameters $g_0 \sim 0.1$ MHz and a single atom cannot couple strongly to the cavity since $g_0 < (\gamma, \kappa)$, where $\gamma$ is the spontaneous decay rate of the excited atomic state and $\kappa$ is the photon loss rate from the cavity. However, the presence of $N_c$ atoms overlapped with the cavity mode increases the effective coupling to $g = g_0\sqrt{N_c}$ and the collective strong coupling between atoms and cavity, defined by $g > (\gamma, \kappa)$, is attained when $N_c$ exceeds a critical number (~$10^4$ in our case). We typically operate our experiment such that $N_c \sim 10^5$.

In our experiments, we keep the cavity tuned to the $F = 2 \to F' = 3$ atomic transition and the ultracold atoms are trapped in the $F = 2$ state. The *probe* beam is derived from an independent ECDL and its frequency $\omega_p/2\pi$ is measured using a saturated absorption spectroscopy (SAS) set up. The *probe* power is controlled using an acousto-optic modulator (AOM) in double-pass configuration. For the VRS measurements, the *probe* beam is coupled into the cavity and its frequency is scanned across the $F = 2 \to F' = 3$ transition while its transmission is monitored on a PMT. With a weak *probe* beam ($P_i \sim 4$ μW) we observe two VRS peaks separated in frequency by $2g$ ($\approx 41 \pm 1$ MHz) in the cavity transmission (Fig. 2). This establishes that system is in the collective strong coupling regime.

In the weak *probe* regime, i.e. low excitation regime, the atom-cavity system behaves like two coupled harmonic oscillators whose degeneracy is lifted by the coupling and the system can described by the extension of the Jaynes-Cummings model [28], the Travis-Cummings Hamiltonian [29]. On increasing the *probe* power, atomic saturation effects come into play since the atom is a two-level system as opposed to a harmonic oscillator with infinite numbers of equally spaced energy levels and so the simple description of coupled harmonic oscillators is not strictly valid [21,30,31]. The other parameter, apart from $g_0$, that determines the behavior of the atom-cavity system is the power of the *probe* beam with respect to the saturation photon number $n_0 = (\gamma_\perp \gamma_\parallel / 4g_0^2) b \approx 48430$ [31], where $\gamma_\parallel = 2\gamma_\perp = 1/\tau$, $b = 8/3$ for a Gaussian standing wave and $\tau = 26.235$ ns is the radiative lifetime of excited state. This value of $n_0$ corresponds to around 60 nW of intra-cavity *probe* power. We study the dependence of VRS on *probe* power as shown in Fig. 2.

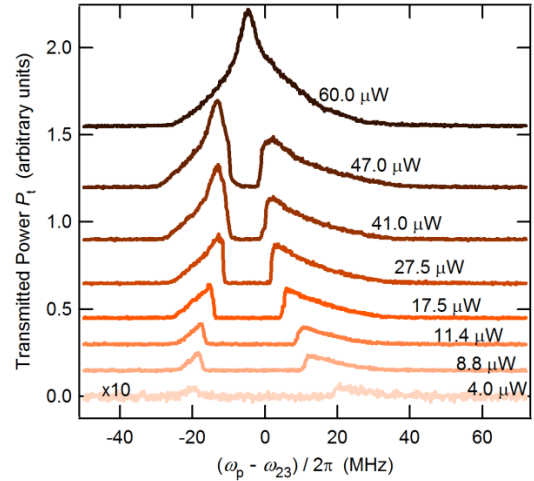

FIG. 2. The transmission spectrum of the coupled atom-cavity system for different incident *probe* laser power $P_i$ (the plots are shifted vertically for clarity). Two clear vacuum-Rabi peaks at low power ($P_i \sim 4$ μW) merge into a single peak as $P_i$ (i.e. atomic excitation) increases. The *probe* laser frequency is scanned from high frequency to low frequency i.e. from right to left in the figure. The asymmetry of the peaks about $\omega_p - \omega_{23} = 0$ is due to OB and small uncontrolled detuning of the cavity from $\omega_{23}/2\pi$.

With increasing *probe* power, the excitation increases and so does the anharmonicity. Two well separated vacuum-Rabi peaks come closer with increasing *probe* power and eventually merge into a single peak [30,31]. The peaks themselves are asymmetric due to an underlying optical bi-stability (see below). The single peak structure at high intensity approaches that of an empty cavity and stems from the $F = 2 \rightarrow F' = 3$ transition being saturated (i.e. leaving fewer atoms in the $F = 2$ level) and thus no longer affecting the cavity transmission significantly. In order to gain additional insight we look at the transmitted power ($P_t$) vs. the incident power ($P_i$) curve of the composite system, traditionally called the OB curve [8].

For the OB curve measurement, we lock the *probe* laser on the $F = 2 \rightarrow F' = 3$ transition and then scan the *probe* power up and down using the AOM. The result is shown in Fig. 3 (filled circles). During the upward-scan the transmitted power ($P_t$) increases until $P_{i,u}$ and then suddenly jumps to a high value, post which $P_t$ again increases with $P_i$. On reversing the scan, i.e. reducing $P_i$, $P_t$ decreases along a different route from the upward-scan until $P_{i,d}$ ($< P_{i,u}$) when it suddenly jumps to a very small value. On further reduction of $P_i$, $P_t$ retraces the upward-scan values. The strongly coupled atom-cavity system thus shows OB and hysteresis much like vapor-cell based atom-cavity system in the weak coupling regime [32,33]. It is interesting to note that at input power $P_{i,u}$, the transmitted power $P_t$ has three values. This could be due to an underlying optical multi-stability [8] or due to slight (uncontrolled) detuning of the cavity from the $F = 2 \rightarrow F' = 3$ transition. The behavior of OB and the hysteresis loop can be controlled by detuning the cavity from the atomic resonance [34–36] which is not discussed further here.

In order to explore the control of the OB in the two-level system discussed above, we extend the scope by involving a third level, $F = 3$. The $F = 2$ and $F = 3$ levels are separated by 3.035 GHz and direct transitions between these levels may be ignored. The dipole allowed transitions $F = 2 \leftrightarrow F' = 3$ and $F = 3 \leftrightarrow F' = 3$ form a Λ-type three level system. As in the earlier measurement, the *probe* laser is locked to the $F = 2 \rightarrow F' = 3$ transition and the additional laser, called the *control* laser, addresses the $F = 3 \leftrightarrow F' = 3$ transition. The *control* laser is derived by frequency shifting a part of the cooling laser beam using an AOM. The same AOM is also used to switch the *control* beam on and off. The *control* beam frequency is -18 MHz (red) detuned from the $F = 3$ and $F' = 3$ transition and its beam diameter is ~ 1 mm. This detuning of the *control* laser is chosen so that it is resonantly couples the $F = 3$ level with the lower frequency vacuum-Rabi peak that appears ~ -20 MHz (red) detuned from the $F' = 3$ level (It was checked, however, by tuning to ~ +28 MHz that the frequency of *control* beam is not very critical for switching). We scan the *probe* power and measure the OB curve for different powers of the *control* beam (Fig. 3).

The hysteresis loop shifts towards higher input powers with increasing *control* beam power $P_c$. The effect can be understood as follows. In absence of the *control* beam, the input *probe* power at which the transmission suddenly jumps to a high value is a measure of the *probe* power where saturation of the $F = 2 \leftrightarrow F' = 3$ transition strongly affects the atom-cavity system. In presence of the *control* beam, a fraction of the atomic population that had decayed to the $F = 3$ level is brought back to the $F = 2$ level via *control* beam absorption followed by spontaneous emission. This increases the population in the $F = 2$ relative to the no-*control* beam case and an increased *probe* power is required for saturation of the $F = 2 \leftrightarrow F' = 3$ transition which results in shifting of the hysteresis loop towards higher input *probe* power. Another major difference from the no-*control* beam case is that during the upward-scan the transmitted power $P_t$ stays almost at zero until the sudden jump. We exploit these features to implement an all-optical switch, at low input *probe* power (~ 2.4 μW), described below. As is evident from Fig. 3, switching can also be implemented for high input power (> 90 μW). We note that the input-output behavior can be controlled by tuning the cavity detuning, probe detuning and control beam detuning [35], which is not discussed further here.

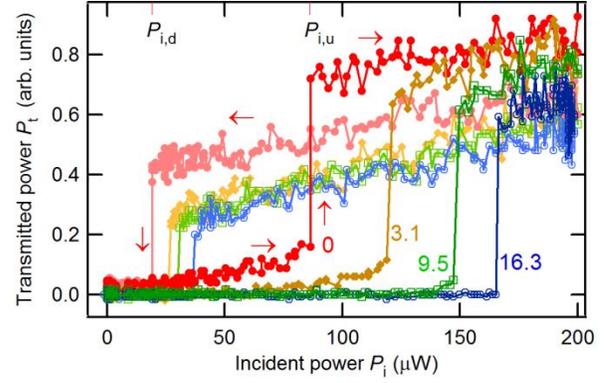

FIG. 3. Transmission of *probe* laser on resonance (i.e. $\omega_p = \omega_{23}$) as the incident *probe* power $P_i$ is varied. Data for different *control* laser power $P_c$ (values in μW indicated) are represented with different symbols. The intensity up-scan (down-scan) data is shown with dark (light) symbols. For clarity, arrows are additionally used to indicate the direction of the intensity scan for $P_c = 0$. The hysteresis shifts towards higher $P_i$ with increasing $P_c$.

Before demonstrating all-optical switching we check the VRS in presence of the *control* beam. For this, the experimental protocol for Fig. 2 is repeated in absence and in presence of the *control* beam. The dotted line in Fig. 4(a) shows the experimentally observed VRS (with $P_i$ ~ 2.4 μW) when the *control* beam is blocked, while the solid line in Fig. 4(a) shows the case when a *control* beam with 9.5 μW optical power is also present. Clearly, the vacuum-Rabi peaks disappear in presence of the *control* beam. This behavior is qualitatively different from that reported by Wei *et al.* [10] who observed suppression of cavity output in a narrow frequency range (~ 5 MHz) within one of the vacuum-Rabi peaks. As discussed in their paper, the optical-switching time observed was fundamentally limited by the inverse of the suppression window. The disappearance of VRS peaks that we observe is not restricted to any frequency window, occurs at ~ 25 times lower *control* beam intensity and is more robust. This allows us to perform robust all-optical switching of the cavity output. We also note that the *control* power required in our experiment is many orders of magnitude lower than in vapor cell based experiments [7]. An efficient all-optical switch should be fast and operate with minimal power, which we demonstrate as follows.

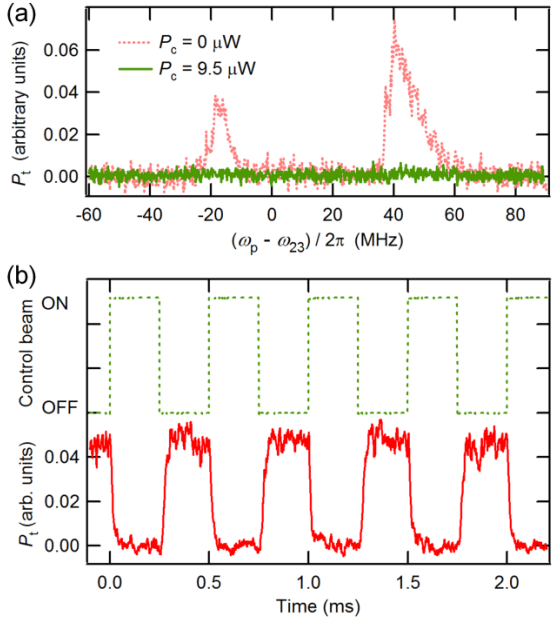

FIG. 4. (a) Transmission spectrum of the coupled atom-cavity system with (solid line) and without (dotted line) the *control* beam. In presence of the *control* beam ($P_c$ = 9.5 μW), the transmission is extinguished over the entire frequency range. (b) *Probe* transmission (solid line) on vacuum-Rabi resonance (i.e. $\omega_p - \omega_{23}$ = -18 MHz) when a periodic train of *control* laser pulses (dotted line) is applied. The *probe* transmission is high when *control* laser is off and low when the *control* laser in on. The switching off time is ~ 12 μs.

The frequency of the *probe* laser is stabilized at -18 MHz (red) detuned from the $F = 2 \leftrightarrow F' = 3$ transition. This detuning of the *probe* laser is chosen because the lower frequency vacuum-Rabi peak has maximum transmission at this detuning of the *probe* laser. Figure 4(b) shows the cavity output power when the *control* beam is switched on and off using an AOM. When the *control* beam is off, the cavity output power is high and steady. On turning the *control* beam on, the cavity output power drops almost to zero. The switching off time $\tau_{off}$, defined as the time required for the output to drop to $1/e$ of the initial value, is 12.5±1.0 μs, 17.4±2.3 μs and 28.0±2.2 μs for $P_c$ = 9.5 μW, 6.1 μW and 3.1 μW, respectively. We see that the cavity output can be switched on and off in micro-second time scales using micro-Watt power levels, i.e. with ~100 pico-Joule of energy. It is expected that the switching times can be reduced by increasing the *control* power but increasing the *control* power beyond 20 μW adversely affects the operation of the DMOT in the current setup.

The relatively simple experimental set-up using a DMOT combines the advantages/abilities of simple vapor cell based atom-cavity systems with those of sophisticated cavity experiments using single atoms. The DMOT holds a steady number of atoms with negligible velocities, prepared in a specific quantum state – this mimics a vapor cell except that complications from thermal motion and mixed quantum states are minimized. This allows continuous measurements and rapid all-optical switching in a different regime, e.g. compared to Sharma *et al.* [32]. At low input *probe* power, the system is in the collective strong-coupling regime. As the input *probe* power is increased, the system makes a transition to the weak coupling regime – this allows flexibility in the operation of a switch. Further manipulation is possible by an off-axis *control* beam which allows control over cavity transmission in different *probe* power regimes and enables rapid all-optical switching. Apart from the conventional applications as an all-optical switch, the possible application of this highly versatile and nonlinear system would be as a very sensitive tool for the measurement of perturbations/interactions, under carefully controlled conditions near the bi-stable region of the system.

In conclusion, we demonstrate that continuous measurements of a strongly coupled atom-cavity system, made possible through a DMOT, results in a simple yet powerful system for cavity QED experiments. We observe OB of the two-level system whose behavior we can control by optical coupling to a third level. Finally, we show all-optical switching of cavity transmission in micro-second timescales with only micro-Watt powers. Future experimental directions include lowering the switching time and the required switching power, and also studies on detuned atom-cavity systems. The rapid measurement demonstrated here also enables the possibility of rapid and non-destructive detection of two-particle interactions [37].

S.D. acknowledges support from Department of Science and Technology (DST), India in form of the DST-INSPIRE Faculty Award (Award No. IFA14-PH-114).